\documentclass[12pt,letterpaper]{article}
\usepackage{amsmath,amssymb,array,calc,rotating,epsfig,psfrag,amscd, cite}

\numberwithin{equation}{section}

\makeatletter
\renewcommand\section{\@startsection {section}{1}{\z@}%
                                   {-3.5ex \@plus -1ex \@minus -.2ex}
                                   {2.3ex \@plus.2ex}%
                                   {\normalfont\large\bfseries}}
\renewcommand\subsection{\@startsection{subsection}{2}{\z@}%
                                     {-3.25ex\@plus -1ex \@minus -.2ex}%
                                     {1.5ex \@plus .2ex}%
                                     {\normalfont\bfseries}}
\makeatother

\let\non\nonumber

\let\S=\Sigma

\newcommand{\bea}{\begin{eqnarray}}
\newcommand{\eea}{\end{eqnarray}}
\newcommand{\be}{\begin{equation}}
\newcommand{\ee}{\end{equation}}

\newcommand{\m}{\mu}

\newcommand{\p}{\partial}

\newcommand{\C}[1]{$(\ref{#1})$}

\def\IZ{\relax\ifmmode\mathchoice
{\hbox{\cmss Z\kern-.4em Z}}{\hbox{\cmss Z\kern-.4em Z}}
{\lower.9pt\hbox{\cmsss Z\kern-.4em Z}} {\lower1.2pt\hbox{\cmsss
Z\kern-.4em Z}}\else{\cmss Z\kern-.4em Z}\fi}
\def\IR{\relax{\rm I\kern-.18em R}}

\def\one{{\hbox{ 1\kern-.8mm l}}}

\newlength{\bredde}
\def\slash#1{\settowidth{\bredde}{$#1$}\ifmmode\,\raisebox{.15ex}{/}
\hspace*{-\bredde} #1\else$\,\raisebox{.15ex}{/}\hspace*{-\bredde}
#1$\fi}

\newsavebox{\zzzbar}
\sbox{\zzzbar}
  {\setlength{\unitlength}{0.9em}
  \begin{picture}(0.6,0.7)
  \thinlines
  \put(0,0){\line(1,0){0.6}}
  \put(0,0.75){\line(1,0){0.575}}
  \multiput(0,0)(0.0125,0.025){30}{\rule{0.3pt}{0.3pt}}
  \multiput(0.2,0)(0.0125,0.025){30}{\rule{0.3pt}{0.3pt}}
  \put(0,0.75){\line(0,-1){0.15}}
  \put(0.015,0.75){\line(0,-1){0.1}}
  \put(0.03,0.75){\line(0,-1){0.075}}
  \put(0.045,0.75){\line(0,-1){0.05}}
  \put(0.05,0.75){\line(0,-1){0.025}}
  \put(0.6,0){\line(0,1){0.15}}
  \put(0.585,0){\line(0,1){0.1}}
  \put(0.57,0){\line(0,1){0.075}}
  \put(0.555,0){\line(0,1){0.05}}
  \put(0.55,0){\line(0,1){0.025}}
  \end{picture}}

\newcommand{\ena}{\end{eqnarray}}
\newcommand{\beqa}{\begin{eqnarray}}
\newcommand{\eeqa}{\end{eqnarray}}

\def\m{\mu}

\def\S{\Sigma}

\begin{document}
\begin{titlepage}

\begin{center}



\vskip 2 cm
{\Large \bf Relations between elliptic modular graphs}\\
\vskip 1.25 cm { Anirban Basu\footnote{email address:
    anirbanbasu@hri.res.in} } \\
{\vskip 0.5cm  Harish--Chandra Research Institute, HBNI, Chhatnag Road, Jhusi,\\
Prayagraj 211019, India}

\end{center}

\vskip 2 cm

\begin{abstract}
\baselineskip=18pt

We consider certain elliptic modular graph functions that arise in the asymptotic expansion around the non--separating node of genus two string invariants that appear in the integrand of the $D^8\mathcal{R}^4$ interaction in the low momentum expansion of the four graviton amplitude in type II superstring theory. These elliptic modular graphs have links given by the Green function, as well its holomorphic and anti--holomorphic derivatives. Using appropriate auxiliary graphs at various intermediate stages of the analysis, we show that each graph can be expressed solely in terms of graphs with links given only by the Green function and not its derivatives. This results in a reduction in the number of basis elements in the space of elliptic modular graphs.

\end{abstract}

\end{titlepage}


\section{Introduction}

S--matrix elements contain very useful information about terms in the low energy effective action of superstring theory. In perturbative string theory, they yield terms with coupling dependence $g_s^{2h-2}$ at genus $h$, where $g_s$ is the string coupling. Each such contribution at fixed $h$ can be expanded in powers of $\alpha'$, the inverse string tension, to yield terms in the effective action that are analytic as well as non--analytic in the external momenta. Focussing on the terms that are analytic in the external momenta, thus we see that the coefficients of the various interactions in the effective action are given by integrals over the moduli space of (super)Riemann surfaces with punctures. These have been explicitly evaluated in various cases for low genera. In such cases, the integrands which depend on the particular string theory under consideration, are given by a sum of terms, each of which has basic building blocks which are referred to as modular graph forms~\cite{DHoker:2015gmr,DHoker:2015wxz}. These are $Sp(2h,\mathbb{Z})$ covariant objects which can be interpreted graphically very usefully. The links of these graphs are given by the conformally invariant Arakelov Green function on the genus $h$ worldsheet or their holomorphic or anti--holomorphic derivatives, while the vertices correspond to the positions of insertions of vertex operators on the worldsheet which are integrated over. Modular graph functions which are $Sp(2h,\mathbb{Z})$ invariant are special cases of modular graph forms, and will be the relevant objects for us. We shall consider them in the context of type II superstring theory, though the techniques should generalize to yield information in cases with lesser supersymmetry.   

Analysis of these string invariants at genus one that result from the low momentum expansion of one loop amplitudes~\cite{Green:1999pv,Green:2008uj,Richards:2008jg,Green:2013bza,DHoker:2019blr} has yielded detailed information about various relations among them, as well as played a crucial role in performing the integrals over moduli space~\cite{DHoker:2015gmr,DHoker:2015sve,DHoker:2015wxz,Basu:2015ayg,Basu:2016fpd,DHoker:2016mwo,Basu:2016xrt,Basu:2016kli,Basu:2016mmk,DHoker:2016quv,Kleinschmidt:2017ege,Basu:2019idd,DHoker:2019blr,Gerken:2019cxz,Gerken:2020yii,Gerken:2020aju}. Analogous analysis has also been done at genus two~\cite{DHoker:2005jhf,DHoker:2013fcx,DHoker:2014oxd,Pioline:2015qha,Basu:2015dqa,DHoker:2017pvk,DHoker:2018mys,Basu:2018bde,DHoker:2020tcq,DHoker:2020uid} based on the low momentum expansion of the two loop four and five graviton amplitudes~\cite{DHoker:2005vch,Berkovits:2005df,Berkovits:2005ng,DHoker:2020prr}, though much less has been understood compared to the one loop graphs. 

Consider the string invariants that arise in the evaluation of the $D^8\mathcal{R}^4$ and $D^6\mathcal{R}^5$ interactions in the low momentum expansion of the four and five graviton amplitudes respectively, at genus two. The asymptotic expansions of these invariants around the non--separating node yield elliptic modular graph functions~\cite{DHoker:2015wxz,DHoker:2020tcq}, graphs in which two of the vertices are not integrated over the worldsheet. These graphs depend on $\tau$, the complex structure of the resulting torus at the non--separating node as well as on $p_a$ and $p_b$, the locations of the two punctures on the genus two Riemann surface which are connected by a long handle, roughly the inverse length of which is the asymptotic expansion parameter\footnote{The expansion yields a Laurent series in this parameter, with the elliptic graphs arising as coefficients in this series.}. In fact, using translational invariance on the torus, these graphs depend only on the difference $v = p_b- p_a$. Thus the two unintegrated vertices are at $v$ and 0. These generalize the modular graph functions in which all the vertices are integrated over\footnote{When the vertices $v$ and 0 are identified, the elliptic modular graphs reduce to modular graphs.}, and are invariant under the $SL(2,\mathbb{Z})$ transformation
\be \tau \rightarrow \frac{a\tau+b}{c\tau+d}, \quad v \rightarrow \frac{v}{c\tau+d},\ee
where $a,b,c,d \in \mathbb{Z}$ and $ad-bc=1$.

Now most of the elliptic modular graph functions that arise in this analysis have links that are given by the Green function on the toroidal worldsheet. However, the ones that arise from the asymptotic expansion of the genus two graphs   
\be \label{4g}\int_{\S_2^2} \prod_{i=1}^2 d^2 z_i \mathcal{G}(z_1,z_2)^2 \m (z_1)\m(z_2)\ee  
and
\be \label{5g}\int_{\S_2^2} \prod_{i=1}^2 d^2 z_i \mathcal{G}(z_1,z_2)^2 (z_1,\overline{z_2})(z_2,\overline{z_1})\ee
are different from this point of view. In fact, \C{4g} and \C{5g} arise in the low momentum expansion of the four and five graviton amplitude respectively. In these expressions, $\mathcal{G} (z_1,z_2)$ is the genus two Arakelov Green function, and 
\be \m(z) = Y^{-1}_{IJ} \omega_I (z) \overline{\omega_J (z)}, \quad (z,\overline{w}) = Y^{-1}_{IJ} \omega_I (z) \overline{\omega_J (w)},\ee
where $Y^{-1}_{IJ} = (Y^{-1})_{IJ}$ and $\omega_I = \omega_I (z) dz$ is the Abelian differential one form. We have defined $Y_{IJ} = ({\rm Im}\Omega)_{IJ}$, where $\Omega$ is the period matrix. The integrals are over $\S_2$, the genus two worldsheet. 

Along with elliptic graphs with links given by the Green function, the asymptotic expansion of \C{4g} and \C{5g} also yields graphs in which the links are given not only by the Green function, but also by its holomorphic and anti--holomorphic derivatives\footnote{In fact, the two graphs \C{4g} and \C{5g} are related~\cite{DHoker:2020tcq} by a non--trivial algebraic identity which also involves other graphs, none of which contain elliptic graphs in their asymptotic expansion that involve derivatives of the Green function. Thus to consider graphs involving derivatives of the Green function, it is enough to either consider \C{4g} or \C{5g}.}. Thus they yield graphs that are of a qualitatively different kind\footnote{This situation is somewhat analogous to modular graphs that arise in the low momentum expansion of the five graviton amplitude at genus one that yield the $D^8\mathcal{R}^5$ and $D^{10} \mathcal{R}^5$ interactions in the type IIB theory, where some of them involve derivatives of the Green function~\cite{Green:2013bza}. However, all these graphs can be expressed in terms of graphs not involving derivatives of the Green function~\cite{Basu:2016mmk}, hence simplifying the structure of the amplitude.}. Essentially this happens because the graphs \C{4g} and \C{5g} form a closed loop on the worldsheet and this involved structure arises from the structure of the Green function in the degeneration limit. It is interesting to analyze if these graphs can be expressed in terms of those that do not have derivatives of Green functions as their links, or if they are genuinely new graphs which add to the number of basis elements in the space of elliptic modular graphs. We shall show that each of these graphs can be expressed solely in terms of those that do not have derivatives of Green functions as their links, leading to a simplification in the structure of the amplitude. It will be interesting to see to what extent this structure survives for elliptic graphs at higher orders in the low momentum expansion. 

We begin by giving the details of the various elliptic modular graphs as well as the other modular graphs that are relevant to our analysis. We then show that each of the graphs having derivatives of Green functions as their links can be expressed in terms of those that have only Green functions as their links, which is the main result of the paper. This is done by introducing auxiliary graphs~\cite{Basu:2016xrt,Basu:2016kli,Basu:2016mmk} at various intermediate stages of the analysis.       

\section{The elliptic modular graph functions having links involving derivatives of the Green function}

We first begin by setting up notations and conventions characterizing the torus that is relevant for our analysis. Denoting by $z$ the coordinate on the torus, we have that
\be -\frac{1}{2} \leq {\rm Re} z \leq \frac{1}{2} , \quad 0 \leq {\rm Im} z \leq \tau_2,\ee
where $\tau = \tau_1 + i\tau_2$ is its complex structure. The measure in the various integrals below is given by $d^2 z= d{\rm Re}z d{\rm Im} z$. In the integrals over the toroidal worldsheet, we shall denote the worldsheet by $\S$.

In the various graphs, the links involve the scalar Green function $G(z,w) = G(z-w)$ on the toroidal worldsheet, as well as their holomorphic and anti--holomorphic derivatives. The Green function is given by the lattice sum~\cite{Lerche:1987qk,Green:1999pv} 
\be \label{Green} G(z) = \frac{1}{\pi} \sum_{(m,n) \neq (0,0)} \frac{\tau_2}{\vert m\tau+n\vert^2}e^{\pi[\bar{z}(m\tau+n)- z(m\bar\tau+n)]/\tau_2}.\ee
It is modular invariant and doubly periodic on the torus. Since it is single valued, we can freely integrate by parts and neglect total derivatives which is very useful in our calculations. 

Also from the definition \C{Green} of the Green function, it follows that
\be \label{vanish}\int_{\S} d^2 z G(z,w)=0.\ee
Thus there can be no graphs where a single link ends on an integrated vertex, a relation which we often use. 

The Green function satisfies the equations
\bea \label{eigen}\overline\p_w\p_z G(z,w) &=& \pi \delta^2 (z-w) - \frac{\pi}{\tau_2}, \non \\
\overline\p_z\p_z G(z,w) &=& -\pi \delta^2 (z-w) + \frac{\pi}{\tau_2},\eea
where the Dirac delta function is normalized such that $\int_{\S}d^2 z \delta^2 (z)=1$.

We now list the elliptic modular graph functions that not only have Green functions as their links, but also its holomorphic and anti--holomorphic derivatives~\cite{DHoker:2018mys}. Apart from the complex structure $\tau$, they are also functions of the complex parameter $v= p_b - p_a$, where $p_a$ and $p_b$ are the locations of the punctures on the genus two worldsheet which are connected by a long handle at the non--separating node.   

To start with, we define
\bea \label{kone}\mathcal{K}_1 (v) = \frac{\tau_2^2}{\pi^2} \int_{\S^2} \frac{d^2 z}{\tau_2} \frac{d^2 w}{\tau_2} \p_z G(z,p_a) \overline\p_z G(z,p_b) G(z,w)^2 \p_w G(w,p_a) \overline\p_w G(w,p_b),\eea
as well as
\bea \label{ktwo}\mathcal{K}_2 (v) = \frac{\tau_2^2}{\pi^2} \int_{\S^2} \frac{d^2 z}{\tau_2} \frac{d^2 w}{\tau_2} \p_z G(z,p_a) \overline\p_z G(z,p_b) G(z,w)^2 \p_w G(w,p_b) \overline\p_w G(w,p_a) \eea
which has the same topology as $\mathcal{K}_1$ but has two derivatives interchanged. Both these graphs are three loop graphs on the toroidal worldsheet\footnote{From now onwards loops refer to the loops in the graph, and not to string loops. All the graphs are at genus one.}.

Apart from these graphs, we also have other multiloop graphs. One of them is defined by
\bea \label{kthree}\widetilde{\mathcal{K}}_3 (v) = \frac{\tau_2^2}{\pi^2} \int_{\S^2} \frac{d^2 z}{\tau_2} \frac{d^2 w}{\tau_2} \Big\vert \p_z G(z,p_a)\Big\vert^2 \p_w G(w,p_a) \overline\p_w G(w,p_b) \Big[ G(z,w)^2 - G(w,p_a)^2\Big].\eea
This graph is complex and in the amplitude we thus have to consider $\widetilde{\mathcal{K}}_3 (v) + c.c.$.  

The remaining graphs are defined by
\bea \label{kfour}\widetilde{\mathcal{K}}_4 (v) = \frac{\tau_2^2}{\pi^2} \int_{\S^2} \frac{d^2 z}{\tau_2} \frac{d^2 w}{\tau_2}\Big\vert \p_z G(z,p_a)\Big\vert^2 \Big\vert \p_w G(w,p_b)\Big\vert^2\Big[G(z,w)^2 -G(w,p_a)^2 \non \\ - G(z,p_b)^2 + G(p_a,p_b)^2\Big],\eea
and 
\bea \label{kfive}\widetilde{\mathcal{K}}_5 = \frac{\tau_2}{\pi}\int_{\S} \frac{d^2 z}{\tau_2} \Big\vert \p_z G(z)\Big\vert^2 \Big[W(z) -4\zeta(3)\Big],\eea
where
\be W(z) = \frac{\tau_2}{\pi} \int_{\S} \frac{d^2 w}{\tau_2} \Big\vert \p_w G(w) \Big\vert^2 \Big(G(z,w)- G(z)\Big)\Big(G(z,w) - G(w)\Big).\ee
Note that $\widetilde{\mathcal{K}}_5$ is independent of $v$, and hence is a modular graph, but not an elliptic modular graph. 

For the sake of brevity, from now onwards, we shall simply refer to all graphs as modular graphs.   

In analyzing these graphs, at certain intermediate steps we obtain expressions involving $G(z,z)$, the Green function at coincident points. We simply set $G(z,z)=0$ in such situations, which is the natural and standard thing to do\footnote{See~\cite{Green:2013bza} for example, for a recent discussion.}. This is because in defining the graphs the vertex operators are never allowed to coincide, as such colliding vertex operators produce another local vertex operator at that point using the operator product expansion. Hence coincident Green functions are not in the moduli space of these graphs, and can be consistently ignored. 

Thus we have that         
\be \int_{\S} d^2 z \Big\vert \p_z G (z,w) \Big\vert^2 =0\ee
where we have integrated by parts, and used \C{eigen} and \C{vanish}. This allows us the simplify the expressions for the various graphs we have listed above.
  
We see that $\mathcal{K}_1 (v)$ and $\mathcal{K}_2 (v)$ are as given in \C{kone} and \C{ktwo} respectively. 
On the other hand, $\widetilde{\mathcal{K}}_3 (v)$ in \C{kthree} reduces to the three loop graph
\bea \label{newthree}\mathcal{K}_3 (v)= \frac{\tau_2^2}{\pi^2} \int_{\S^2} \frac{d^2 z}{\tau_2} \frac{d^2 w}{\tau_2}\Big\vert \p_z G(z,p_a)\Big\vert^2 \p_w G(w,p_a) \overline\p_w G(w,p_b) G(z,w)^2 ,\eea
and so we shall consider $\mathcal{K}_3 (v) + c.c.$.

Also $\widetilde{\mathcal{K}}_4 (v)$ in \C{kfour} reduces to a three loop graph
\bea \label{newfour}\mathcal{K}_4 (v)= \frac{\tau_2^2}{\pi^2} \int_{\S^2} \frac{d^2 z}{\tau_2} \frac{d^2 w}{\tau_2}\Big\vert \p_z G(z,p_a)\Big\vert^2 \Big\vert \p_w G(w,p_b)\Big\vert^2 G(z,w)^2 \eea
with distinct topology compared to $\mathcal{K}_3 (v)$.

Finally, $\widetilde{\mathcal{K}}_5$ in \C{kfive} reduces to a four loop graph given by
\bea \label{newfive}\mathcal{K}_5 = \frac{\tau_2^2}{\pi^2}  \int_{\S^2} \frac{d^2 z}{\tau_2} \frac{d^2 w}{\tau_2} \Big\vert \p_z G(z)\Big\vert^2  \Big\vert \p_w G(w) \Big\vert^2 \Big(G(z,w)- G(z)\Big)\Big(G(z,w) - G(w)\Big) ,\eea
since only the contribution involving $W(z)$ survives. Thus among the graphs we consider, this is the only graph with more than one term. We shall depict all these graphs diagrammatically later on.

\begin{figure}[ht] 
\begin{center}
\[
\mbox{\begin{picture}(250,90)(0,0)
\includegraphics[scale=.7]{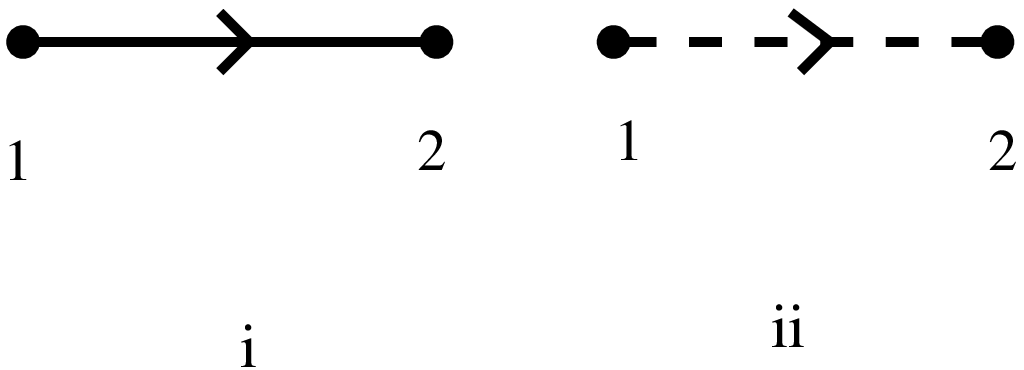}
\end{picture}}
\]
\caption{(i) $\p_{z_2} G(z_1,z_2) = -\p_{z_1} G(z_1,z_2)$,  (ii) $\overline\p_{z_2} G(z_1,z_2) = -\overline\p_{z_1} G(z_1,z_2)$}
\end{center}
\end{figure}

We shall need the expression for the $SL(2,\mathbb{Z})$ invariant Laplacian involving derivatives with respect to the complex structure given by
\be \Delta_\tau = 4\tau_2^2\frac{\p^2}{\p\tau \p\overline\tau}.\ee

\begin{figure}[ht]
\begin{center}
\[
\mbox{\begin{picture}(130,45)(0,0)
\includegraphics[scale=.6]{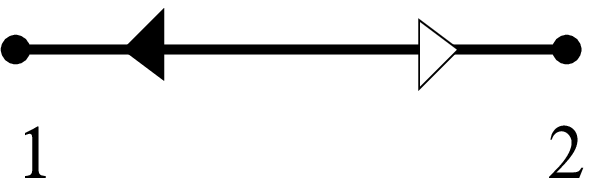}
\end{picture}}
\]
\caption{$\overline\p_{z_1} \p_{z_2} G(z_1,z_2)$}
\end{center}
\end{figure}

We shall also find useful the $SL(2,\mathbb{Z})$ invariant Laplacian involving derivatives with respect to the complex parameter $v$ on the torus given by  
\be \Delta_v = 4\tau_2\frac{\p^2}{\p v \p \overline{v}}.\ee

 In the various graphs, the unintegrated vertices $v$ and $0$ can be interchanged which we shall use implicitly, which simply follows from the properties of the Green function \C{Green}. 

In our analysis, it is very useful to diagrammatically depict these graphs. While the Green function along a link is given by a solid line, holomorphic and anti--holomorphic derivatives of Green functions are depicted by figure 1. Such derivatives appearing in the same link is depicted by figure 2.

\section{Relations between the various modular graphs}

\begin{figure}[ht]
\begin{center}
\[
\mbox{\begin{picture}(360,270)(0,0)
\includegraphics[scale=.7]{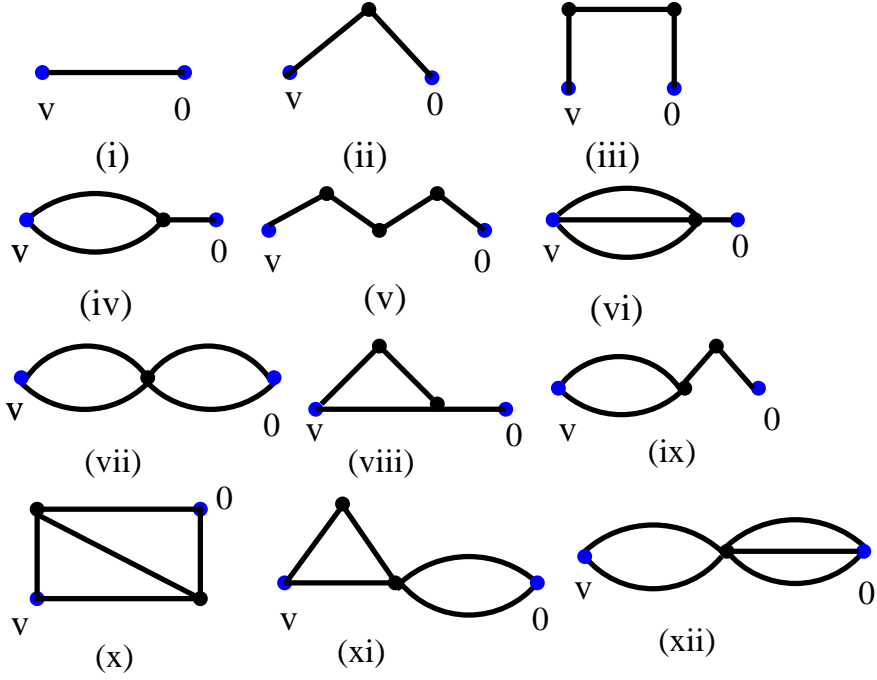}
\end{picture}}
\]
\caption{The elliptic modular graphs (i) $G (v)$, (ii) $G_2 (v)$, (iii) $G_3 (v)$, (iv) $D_3^{(1)} (v)$, (v) $G_4 (v)$, (vi) $D_4^{(1)} (v)$, (vii) $D_4^{(2)} (v)$, (viii) $D_4^{(1,2)} (v)$, (ix) $D_4^{(1,1,2)} (v)$, (x) $D_5^{(2,1,2)} (v)$, (xi) $D_5^{(1,2,2)} (v)$, (xii) $D_5^{(2)} (v)$}  
\end{center}
\end{figure}

Apart from the graphs $\mathcal{K}_i (v)$ ($i=1,\ldots,5$) mentioned above, several other graphs appear in our analysis which we now list. The relevant elliptic modular graphs are given in figure 3. The unintegrated vertices of such graphs (denoted by $v$ and 0) are always depicted in the figures, while all other vertices are integrated over. 

From now onwards, for the sake of brevity we shall refer to $\int_{\S} d^2 z/\tau_2$ simply as $\int_{z}$ in the various expressions.   

Some of the graphs in figure 3 are defined by~\cite{DHoker:2018mys}
\be D_l^{(k)} (v) = \int_z G^{k} (v,z)G (z)^{l-k} = D_l^{(l-k)} (v)\ee
for $l \geq 2$ and $k \leq l$. Some of the others are given by the iterated Green functions $G_k (v)$ ($k \geq 1$) defined recursively by
\be G_{k+1} (v) = \int_z G(v,z) G_k (z)\ee
where $G_1 (z) = G (z)$, the Green function.  The remaining graphs are defined by the integrals
\bea D_4^{(1,2)} (v) &=& \int_{wz} G(w) G(w,v) G(z,v) G(w,z), \non \\ D_4^{(1,1,2)} (v) &=& \int_{wz} G(w) G(z,v)^2 G(w,z), \non \eea
\bea D_5^{(2,1,2)} (v) &=& \int_{wz} G(w)G(z) G(w,v) G(z,v) G(w,z), \non \\ D_5^{(1,2,2)} (v) &=& \int_{wz} G(w)^2 G(w,v) G(z,v) G(w,z).\eea

We next mention the various modular graphs that arise in our analysis which are given in figure 4.  

\begin{figure}[ht]
\begin{center}
\[
\mbox{\begin{picture}(280,95)(0,0)
\includegraphics[scale=.65]{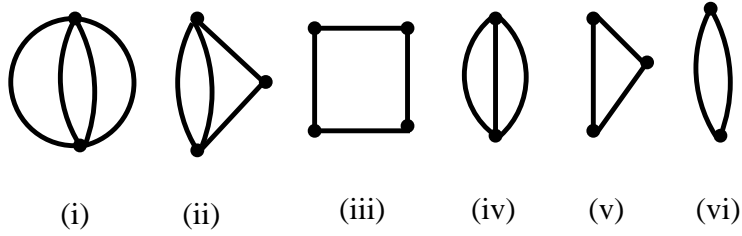}
\end{picture}}
\]
\caption{(i) $D_4$, (ii) $C_{1,1,2}$, (iii) $E_4$, (iv) $D_3$, (v) $E_3$, (vi) $E_2$}
\end{center}
\end{figure}

These can be obtained from some of the elliptic modular graphs mentioned above by identifying $v$ and 0. Thus we have that $D_l^{(k)} (0) = D_l$ for $l \geq 2, k \leq l$, and $G_{k+1} (0) = E_{k+1}$ for $k \geq 1$. We also have that
\be D_4^{(1,2)} (0) = D_4^{(1,1,2)} (0) = C_{1,1,2}.\ee

\begin{figure}[ht]
\begin{center}
\[
\mbox{\begin{picture}(260,28)(0,0)
\includegraphics[scale=.6]{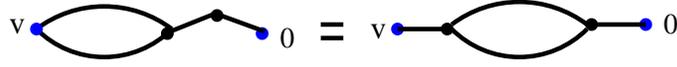}
\end{picture}}
\]
\caption{A relation between elliptic graphs}
\end{center}
\end{figure}

We also often use the relation between the elliptic graphs given in figure 5~\cite{Basu:2020pey} , where the graph on the right hand side is given by
\be \int_{wz} G(z,v) G(w) G(w,z)^2.\ee

We now proceed to obtain the relation between each of the graphs $\mathcal{K}_i (v)$ ($i=1,\ldots,5$) and the other graphs given in figures 3 and 4. Note that all the links in all the graphs in figures 3 and 4 involve only the Green function, and not its derivatives.   

Thus schematically the aim is to express each of the graphs $\mathcal{K}_i (v)$ in terms of graphs having no derivatives of Green functions as the links. The strategy we shall follow starting from $\mathcal{K}_i(v)$ is to use the single valuedness of the Green function \C{Green} and integrate by parts, and use \C{eigen} to remove the derivatives. Now this does not always work in a straightforward way in removing derivatives, as will be evident in the manipulations below. To proceed in such cases, we shall introduce auxiliary graphs at various intermediate stages of the analysis. To summarize, at the end we shall express each $\mathcal{K}_i (v)$ in terms of graphs that appear in figures 3 and 4.    

\subsection{Relation involving $\mathcal{K}_1 (v)$}

We first consider the relation involving the graph $\mathcal{K}_1 (v)$ defined by \C{kone}.

\begin{figure}[ht]
\begin{center}
\[
\mbox{\begin{picture}(140,80)(0,0)
\includegraphics[scale=.7]{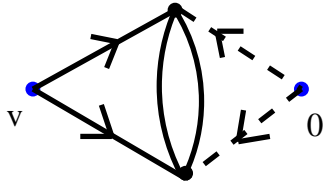}
\end{picture}}
\]
\caption{$\pi^2 \mathcal{K}_1 (v)$}  
\end{center}
\end{figure}

Thus we have that
\bea \mathcal{K}_1 (v) = \frac{\tau_2^2}{\pi^2} \int_{wz} \p_z G(z,v) \overline\p_z G(z) G(z,w)^2 \p_w G(w,v) \overline\p_w G(w)\eea
as depicted by figure 6.

Integrating appropriately by parts and using \C{eigen}, we get that
\bea \label{intk1}\mathcal{K}_1 (v) = \frac{4}{3} D_3 G(v) - \frac{G(v)^4}{3} - D_4^{(1,1,2)} (v) -2 F_1 (v) -  \Big(F_2 (v) + c.c.\Big) +2H_1 (v),\eea
where
\bea H_1 (v) = \frac{\tau_2^2}{\pi^2} \int_{wz}  \p_z G(z,w) \overline\p_w G(z,w) \p_z G(z,v) \overline\p_w G(w) G(w,v) G(z),\eea
as depicted in figure 7. The graphs $F_1 (v)$ and $F_2 (v)$ are defined by \C{defF} and given in figure 13. 

\begin{figure}[ht]
\begin{center}
\[
\mbox{\begin{picture}(280,320)(0,0)
\includegraphics[scale=.65]{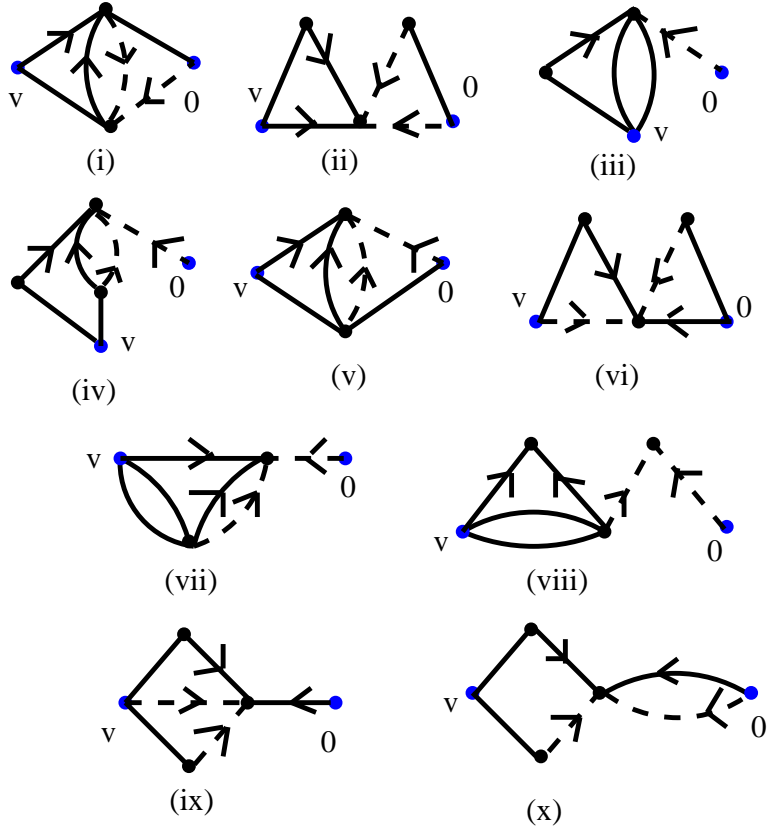}
\end{picture}}
\]
\caption{The graphs (i) $\pi^2 H_1 (v)$, (ii) $\pi^2 H_2 (v)$, (iii) $\pi H_3 (v)$, (iv) $\pi^2 H_4 (v)$, (v) $\pi^2 H_5 (v)$ , (vi) $\pi^2 H_6 (v)$, (vii) $\pi^2 H_7 (v)$, (viii) $\pi^2 H_8 (v)$, (ix) $\pi^2 H_9 (v)$, (x) $\pi^2 H_{10} (v)$}  
\end{center}
\end{figure}

Thus using \C{relF}, we see that all the terms on the right hand side of \C{intk1} except the one involving $H_1 (v)$ are expressible in terms of graphs having only the Green function as the links.  

Now let us consider the term $H_1 (v)$. To analyze it, we shall introduce appropriate auxiliary graphs~\cite{Basu:2016xrt,Basu:2016kli,Basu:2016mmk}. In general, an auxiliary graph is a graph which trivially yields the desired graph using \C{eigen}. However, it can be evaluated independently by integrating by parts and using \C{eigen} such that all the derivatives are removed. Thus equating the two distinct ways of evaluating the auxiliary graph, we get an expression in which the derivatives in the desired graph are completely removed, in the sense that it is expressed in terms of graphs without derivatives of the Green function as the links. Often this complete reduction requires introducing more than one auxiliary graph at various intermediate stages of the analysis.         

\begin{figure}[ht]
\begin{center}
\[
\mbox{\begin{picture}(350,335)(0,0)
\includegraphics[scale=.6]{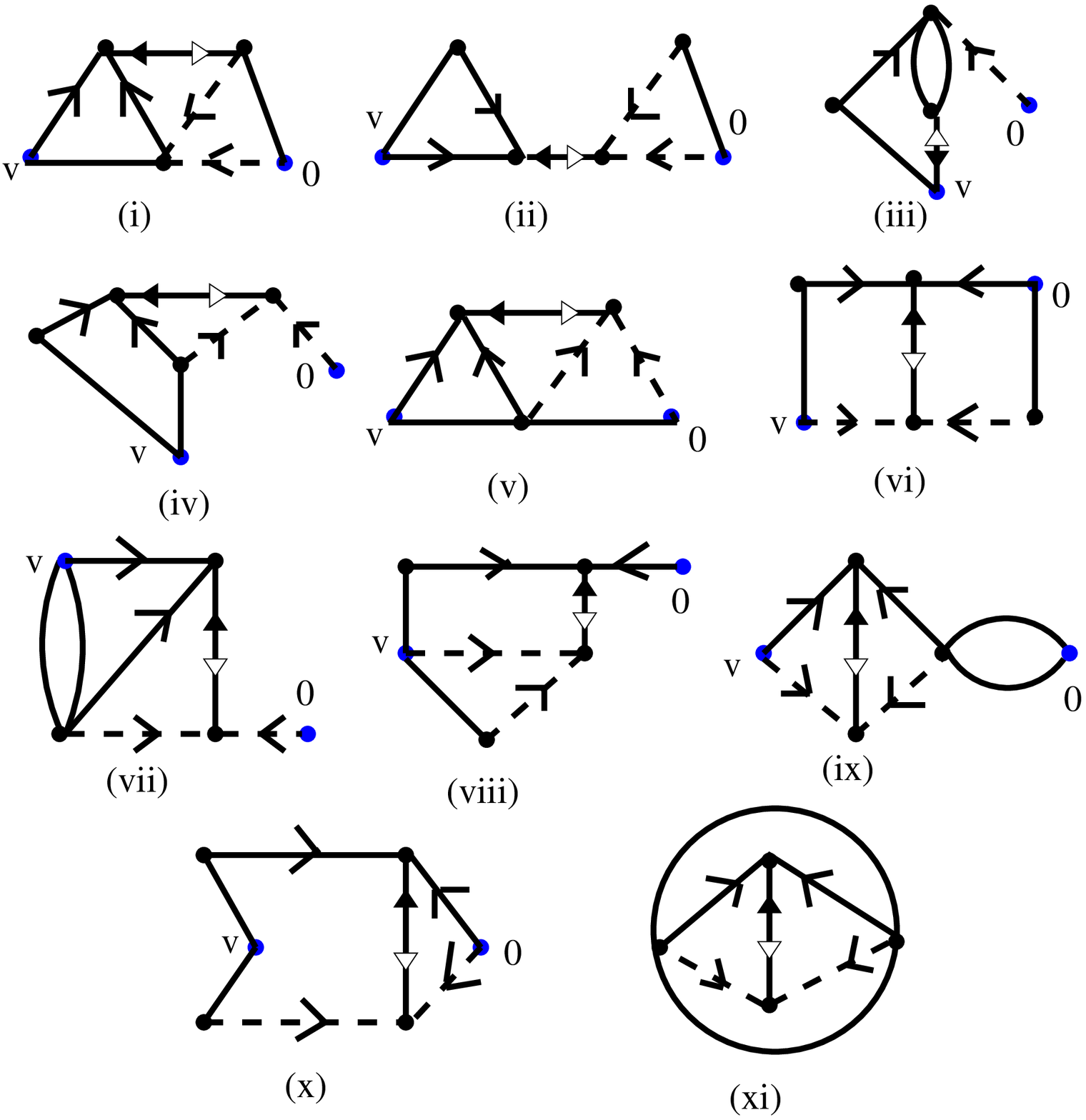}
\end{picture}}
\]
\caption{The auxiliary graphs (i) $\pi^3 A_1  (v)$, (ii) $\pi^3 A_2 (v)$, (iii) $\pi^2 A_3 (v)$, (iv) $\pi^3 A_4 (v)$, (v) $\pi^3 A_5 (v)$, (vi) $\pi^3 A_6 (v)$, (vii) $\pi^3 A_7 (v)$, (viii) $\pi^3 A_8 (v)$, (ix) $\pi^3 A_9 (v)$, (x) $\pi^3 A_{10} (v)$, (xi) $\pi^3 A_{11} (v)$}  
\end{center}
\end{figure}

To analyze $H_1 (v)$, we begin with the auxiliary graph $A_1 (v) + c.c.$, where
\bea A_1 (v) = \frac{\tau_2^3}{\pi^3} \int_{xwz} \p_w\overline\p_z G(z,w) \p_z G(z,v) \p_z G(x,z) \overline\p_x G(x,w) \overline\p_x G(x) G(w) G(x,v), \eea
as given in figure 8. It is evaluated trivially using \C{eigen} for the link having a $\p$ as well as a $\overline\p$ acting on the Green function, and alternatively by moving the derivatives around the circuit and using \C{eigen}. This will be the strategy we shall always follow.  

This leads to
\bea \label{H1}&& H_1 (v) = \frac{D_4^{(1)} (v)}{2} +\frac{E_2^2}{4} + D_4^{(1,2)} (v) -\frac{D_4^{(1,1,2)} (v)}{2} -\frac{D_4^{(2)} (v)}{2} +\frac{G_2(v)^2}{2} +\frac{G(v)^4}{4} \non \\ && +F_1 (v) +\frac{1}{2} \Big(F_3 (v) +c.c.\Big) -\frac{1}{4} \Big(F_4 (v) +c.c.\Big)-\frac{1}{2} \Big(F_5 (v) + c.c.\Big) \non \eea
\bea && -\frac{1}{4} \Big(F_6 (v) + c.c.\Big) +\frac{1}{2}\Big(\frac{\tau_2}{\pi} \p_v G_2 (v) \overline\p_v G_3 (v) + c.c.\Big) +H_2 (v) -\frac{1}{4}\Big(H_3 (v) +c.c.\Big),\non \\ \eea
where we define
\bea H_2 (v) &=& \frac{\tau_2^2}{\pi^2} \int_{wxz} G(z,v) G(w) \p_x G(x,z) \p_x G(x,v) \overline\p_x G(x,w) \overline\p_xG(x), \non \\ H_3 (v) &=& \frac{\tau_2}{\pi} \int_{wz}  G(z,v)^2 G(w,v) \p_z G(w,z) \overline\p_z G(z)\eea
as depicted by figure 7. Also the graphs $F_i (v)$ are defined by \C{defF} and depicted by figure 13. 

From the relations in \C{relF} and \C{reldel}, we see that apart from the terms involving $H_2 (v)$ and $H_3 (v)$ on the right hand side of \C{H1}, all the other terms can be expressed in terms of graphs without derivatives of the Green function as the links. To analyze $H_2 (v)$, we consider the auxiliary graph
\bea A_2 (v) = \frac{\tau_2^3}{\pi^3} \int_{wxyz}  G(z,v)G(w)\overline\p_x \p_y G(x,y) \p_x G(x,z) \p_x G(x,v) \overline\p_y G(y) \overline\p_y G(w,y),\non \\ \eea
as given in figure 8, which leads to
\bea \label{k23}H_2 (v) = Q_1 Q_1^* - D_4^{(1,1,2)} (v) +\frac{D_4^{(2)} (v)}{4} -\frac{E_2^2}{4} + G_4 (v),\eea
where $Q_1$ is defined by \C{defQ}. From \C{relQ} we see that $H_2 (v)$ is expressible in terms of graphs without derivatives of the Green function as its links.

Next to analyze $H_3 (v) + c.c.$, we consider the auxiliary graph\footnote{This analysis is similar to that in~\cite{Basu:2019idd}.}
\bea A_3 (v) = \frac{\tau_2^2}{\pi^2} \int_{wxz} G(x,v) G(z,w)^2 \p_z G(x,z) \overline\p_z G(z) \p_w \overline\p_v G(w,v),\eea
depicted by figure 8,
which leads to the relation
\be \label{k25}H_3 (v) = E_2 G_2 (v) - G_2 (v)^2 + E_4 - 2 H_4 (v),\ee
where the graph $H_4 (v)$ is defined by
\bea H_4 (v) = \frac{\tau_2^2}{\pi^2} \int_{wxz} G(x,v) G(w,v) \p_z G(x,z) \p_z G(w,z) \overline\p_z G(w,z) \overline\p_z G(z)\eea
as depicted by figure 7.
To analyze $H_4 (v)$, we consider the auxiliary graph $A_4 (v)$ defined by
\bea A_4 (v) = \frac{\tau_2^3}{\pi^3} \int_{wxyz}  G(y,v) G(x,v) \p_w G(w,y) \p_w G(w,x) \overline\p_z G(x,z) \overline\p_z G(z)  \overline \p_w \p_z G(w,z)\eea
depicted by figure 8. This leads to the relation
\bea \label{k26}H_4 (v) &=& G_4 (v) +\frac{3 E_4}{2}- D_4^{(1,2)} (v) - \frac{D_4^{(2)} (v)}{2} - \frac{E_2^2}{2} +\frac{C_{1,1,2}}{2} - \frac{1}{2}G(v) D_3^{(1)} (v) \non \\ &&- \frac{\tau_2}{\pi} \overline\p_v G_2 (v)\p_v G_3 (v) +\frac{F_2(v)}{2} -  F_3 (v) +\frac{F_6 (v)}{2}+F_7 (v) .  \eea
Using the definitions in \C{defF}, and the relations \C{relF} and \C{reldel} we see that the we have expressed $H_3 (v)$ in terms of graphs without derivatives of the Green function as their links. 

Thus from the various relations above, we see that $\mathcal{K}_1 (v)$ is expressible in terms of the modular graphs given in figures 3 and 4, in which the links are given by the Green function and not its derivatives. In fact, adding the various contributions, we get that 
\bea \label{K1}\mathcal{K}_1 (v) &=& \frac{1}{4}\Big(\Delta_\tau -8\Big)E_2^2 + 2 E_4 + 4 G_4 (v) + G_2 (v)^2 - \frac{D_4^{(2)} (v)}{2} \non \\ &&-2 D_4^{(1,1,2)} (v) -\frac{2}{3} D_3 G(v)+\frac{G(v)^4}{6}.\eea

We now analyze the remaining graphs $\mathcal{K}_i (v)$ where the basic strategy is the same, and hence we shall be somewhat brief. We shall say that a graph has ``simplified'', by which we shall always mean that it is expressible in terms of graphs where the links involve only the Green function.     

\subsection{Relation involving $\mathcal{K}_2 (v)$}

We next consider the graph $\mathcal{K}_2 (v)$ defined by \C{ktwo} leading to
\bea \mathcal{K}_2 (v) = \frac{\tau_2^2}{\pi^2} \int_{wz}  \p_z G(z,v) \overline\p_z G(z) G(z,w)^2 \p_w G(w) \overline\p_w G(w,v), \eea
as depicted in figure 9.

\begin{figure}[ht]
\begin{center}
\[
\mbox{\begin{picture}(140,80)(0,0)
\includegraphics[scale=.7]{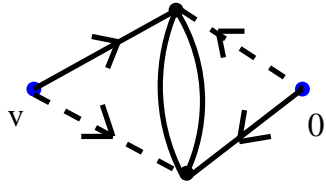}
\end{picture}}
\]
\caption{$\pi^2 \mathcal{K}_2 (v)$}  
\end{center}
\end{figure}

This gives us that
\bea \label{k21}\mathcal{K}_2 (v) &=& \frac{2 G(v)^4}{3} - \frac{2}{3} D_3 G(v) -2D_4^{(1,2)} (v) + 2 G(v) D_3^{(1)}  (v) - D_4^{(2)} (v) \non \\ &&+ G_2(v)^2-\Big(F_2 (v) + c.c.\Big) - \mathcal{K}_1 (v) + 2 H_5(v),\eea
where $H_5 (v)$ is defined by
\bea H_5 (v) = \frac{\tau_2^2}{\pi^2} \int_{wz}  G(w,v) G(w) \p_z G(z,v) \p_z G(z,w) \overline\p_z G(z,w) \overline\p_z G(z)\eea
as given in figure 7. Thus using \C{relF} and \C{K1} we see that apart from the term involving $H_5 (v)$, the right hand side of \C{k21} has simplified. Now to simplify $H_5 (v)$, we start with the auxiliary graph $A_5 (v)$ defined by
\bea A_5 (v) = \frac{\tau_2^3}{\pi^3} \int_{wxz}  G(x) G(x,v) \p_z G(z,v) \p_z G(x,z) \overline\p_w G(w,x) \overline\p_w G(w) \overline\p_z \p_w G(w,z)\eea
as depicted by figure 8.
This leads to the expression
\bea \label{k22}H_5 (v) &=& - \frac{1}{2} G(v) D_3^{(1)}(v) +\frac{2 F_8 (v)}{3} - \frac{1}{2} G(v) F_9 (v) -\frac{D_4^{(1)}(v)}{3}  + 2 D_4^{(1,2)} (v) \non \\ && + E_2 G_2 (v) + G_2(v)^2 - D_4^{(2)}(v)+ \Big(F_3 (v) + c.c.\Big) \non \\&& +\frac{1}{2}\Big(H_3 (v) + c.c.\Big)- \frac{1}{2}\Big(\frac{\tau_2}{\pi} \p_v G_2 (v) \overline\p_v D_3^{(1)} (v)+ c.c.\Big)+H_2 (v) + H_6 (v), \non \\ \eea
where
\bea H_6 (v) = \frac{\tau_2^2}{\pi^2} \int_{wxz}  G(z,v) G(w) \p_x G(x,z) \overline\p_x G(x,v) \p_x G(x) \overline\p_x G(w,x)\eea
as depicted by figure 7, and the graphs $F_i (v)$ are defined by \C{defF}. Thus using \C{relF}, \C{reldel} and the expressions involving $H_2 (v)$ and $H_3 (v)$ in the analysis of $\mathcal{K}_1 (v)$, we see that all the terms in the right hand side of \C{k22} have simplified, except the one involving $H_6 (v)$.

To simplify $H_6 (v)$, we consider the auxiliary graph $A_6 (v)$  defined by
\bea A_6 (v)=\frac{\tau_2^3}{\pi^3} \int_{wxyz}  G(x,v) G(y) \p_z G(x,z) \p_z G(z) \overline\p_w G(w,v) \overline\p_w G(w,y)  \overline\p_z \p_w G(w,z),\eea
as given in figure 8, 
which leads to the relation
\bea \label{k24}H_6 (v) &=& Q_2 (v) Q_2^* (v)+ G_4 (v) + F_1 (v) - \Big(F_{10} (v) + c.c.\Big)\non \\ &&+ \Big(\frac{\tau_2}{\pi} \p_v G_2 (v) \overline\p_v D_3^{(1)} (v)+c.c.\Big)  -  \frac{\tau_2}{\pi} G(v) \p_v G_2 (v) \overline\p_v G_2 (v).\eea
Thus using \C{defQ} and the relations \C{relF}, \C{relQ} and \C{reldel}, we see that $H_6 (v)$ has simplified.
Adding the various contributions, we get that
\bea \label{K2}&&\mathcal{K}_2 (v)  = -\frac{1}{4\pi} \Delta_v \Big(D_5^{(2,1,2)} (v) + D_5^{(1,2,2)} (v) - \frac{2D_5^{(2)}(v)}{3} -2 G_2 (v) G_3 (v)  - G_2 (v) D_3^{(1)} (v)\Big)\non \\ &&-\frac{1}{4\pi} G(v)\Delta_v \Big(G_2(v)^2 +\frac{D_4^{(2)}(v)}{2}\Big) +\frac{1}{4} \Big(\Delta_\tau +2\Big)E_2^2  +\frac{1}{4}\Big(\Delta_\tau +12\Big) G_2(v)^2 +10 D_4^{(1,2)} (v)  \non \\ && - \frac{5}{2} D_4^{(2)} (v) - 3 E_2 G_2 (v)  + 5 D_4^{(1,1,2)} (v)    + 6 \Big(G_4 (v) - E_4\Big)+ G(v) \Big( \frac{2G(v)^3}{3} -\frac{8D_3}{3} \non \\&& + D_3^{(1)} (v) - E_2 G(v) + 4 G_3 (v) - 2 E_3 - G(v) G_2 (v)\Big)-\mathcal{K}_1(v)\eea
and hence $\mathcal{K}_2 (v)$ has simplified\footnote{In fact, it is only the combination $\mathcal{K}_1 (v) + \mathcal{K}_2 (v)$ that arises in the asmyptotic expansion of the $D^8\mathcal{R}^4$ string invariants~\cite{DHoker:2018mys}, which is directly given by \C{K2}.}. 

\subsection{Relation involving $\mathcal{K}_3 (v)$}

We next consider the graph $\mathcal{K}_3 (v)$ defined by \C{newthree} which we write as
\bea \mathcal{K}_3 (v)= \frac{\tau_2^2}{\pi^2} \int_{wz} \Big\vert \p_z G(z,v)\Big\vert^2 \p_w G(w,v) \overline\p_w G(w) G(z,w)^2 \eea
as depicted by figure 10.

\begin{figure}[ht]
\begin{center}
\[
\mbox{\begin{picture}(140,70)(0,0)
\includegraphics[scale=.7]{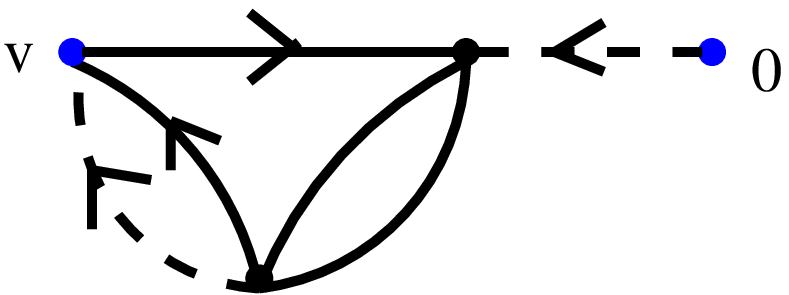}
\end{picture}}
\]
\caption{$\pi^2 \mathcal{K}_3 (v)$}  
\end{center}
\end{figure}

Proceeding to simplify it, we get that
\bea \label{k31}\mathcal{K}_3 (v)  =  D_3 G(v) -  D_4^{(1,1,2)} (v)  -  F_2 (v) -  F_4 (v)  +H_7 (v),\eea
where $H_7 (v)$ is defined by
\bea H_7 (v) =  \frac{\tau_2^2}{\pi^2} \int_{wz} G(w,v)^2 \p_z G(w,z) \p_z G(z,v) \overline\p_z G(w,z) \overline\p_z G(z)\eea
as depicted by figure 7. Using \C{relF} we see that this is the only term on the right hand side of \C{k31} which has not simplified.

To simplify it, we consider the auxiliary graph $A_7 (v)$ defined by
\bea A_7 (v) = \frac{\tau_2^3}{\pi^3} \int_{wxz} G(w,v)^2 \p_z G(z,v) \p_z G(w,z) \overline\p_x G(w,x) \overline\p_x G(x) \overline\p_z \p_x G(x,z)\eea
as given in figure 8. This yields the expression
\bea \label{k32}H_7 (v) &=& \frac{G(v)^4}{12} -\frac{D_4}{12} + C_{1,1,2} + E_2 G_2 (v) + E_2^2 + D_4^{(1,1,2)} (v) - \frac{1}{3} D_3 G(v) \non \\ &&- H_3 (v) -  F_6(v) +\frac{F_8 (v)}{3} - \frac{\tau_2}{\pi} \p_v G_2 (v) \overline\p_v D_3^{(1)} (v) + H_8 (v),\eea
where we define
\bea H_8 (v) = \frac{\tau_2^2}{\pi^2} \int_{wxz}  G(w,v)^2 \p_z G(z,v) \p_z G(w,z) \overline\p_x G(w,x) \overline\p_x G(x)\eea
as given in figure 7. Based on \C{relF}, \C{reldel} and the previous analysis of $H_3 (v)$, this is the only term which has not simplified in the expression for $H_7 (v)$. 

To simplify it, we rewrite it as\footnote{This is obtained by using \C{Belt} as discussed in the appendix.}
\bea \label{k33}&&H_8 (v) + c.c. = \frac{1}{3}\Delta_\tau D_4^{(1)} (v) - 2 C_{1,1,2} + E_2^2 - E_4  -2 F_{11}(v) \non \\ && + \Big(H_3 (v) + c.c.\Big)+\frac{2\tau_2}{\pi} G(v)\p_v G_2 (v) \overline\p_v G_2 (v)- 2\Big(H_9 (v) + c.c.\Big),\eea
where the graph
\bea H_9 (v) = \frac{\tau_2^2}{\pi^2} \int_{wxz}  G(v,z)G(v,w) \p_x G(x,z) \p_x G(x) \overline\p_x G(w,x) \overline\p_x G(v,x)\eea
depicted by figure 7, is the only term that has not simplified in \C{k33} on using \C{relF}, \C{reldel} and the expression for $H_3 (v)$.  

Now to simplify $H_9 (v)$, consider the auxiliary graph
\bea A_8 (v) = \frac{\tau_2^3}{\pi^3} \int_{wxyz} G(v,z) G(v,y) \p_w G(w,z) \p_w G(w)  \overline\p_x G(x,y) \overline\p_x G(x,v) \overline\p_w \p_x G(x,w)\eea
given in figure 8, which leads to
\bea \label{k34} H_9 (v)  &=&  Q_1^* Q_2 (v) +  E_4 + \frac{D_4^{(1)} (v)}{2} - \frac{1}{2}E_2 G_2 (v) - D_4^{(1,2)}(v) -\frac{C_{1,1,2}}{2}-\frac{F_4 (v)}{2} \non \\ &&+F_{10} (v) -\frac{\tau_2}{\pi} \p_v G_2 (v) \overline\p_v G_3 (v) +\frac{\tau_2}{2\pi} \p_v G_2 (v) \overline\p_v D_3^{(1)} (v).\eea
Thus on using \C{relF}, \C{relQ} and \C{reldel} we see $H_9 (v)$ has simplified.

Hence adding the various contributions, we get that\footnote{This rectifies a typo in the earlier versions. I am thankful to Boris Pioline for useful discussions on this issue.}
\bea \label{3k}&&\mathcal{K}_3 (v) + c.c.= \frac{1}{2\pi}\Delta_v \Big(G_2 (v) G_3 (v) - G_2 (v) D_3^{(1)}(v) - \frac{1}{2} D_5^{(1,2,2)} (v) +\frac{1}{6} D_5^{(2)} (v)\Big)\non \\ &&+\frac{1}{4\pi} G(v)\Delta_v G_2(v)^2 +\frac{1}{3}\Big(\Delta_\tau -5\Big) D_4^{(1)}(v)-\frac{1}{2}\Big(\Delta_\tau -14\Big) \Big( E_2 (v) G_2 (v)\Big)\non \\ &&+ 3 G_2(v)^2 + G(v)\Big(4G_3(v)-3 D_3^{(1)} (v)+\frac{D_3}{3}+\frac{G(v)^3}{6} -E_2 G(v) -2 E_3\Big)\non \\&& -\frac{D_4}{6} +3 C_{1,1,2} + 3 E_2^2 - 8 E_4 + 2 D_4^{(1,1,2)}(v) - D_4^{(2)} (v) + 2 G_4(v)\eea
showing that $\mathcal{K}_3 (v) + c.c.$ has simplified. 

\subsection{Relation involving $\mathcal{K}_4 (v)$}

\begin{figure}[ht]
\begin{center}
\[
\mbox{\begin{picture}(210,60)(0,0)
\includegraphics[scale=.8]{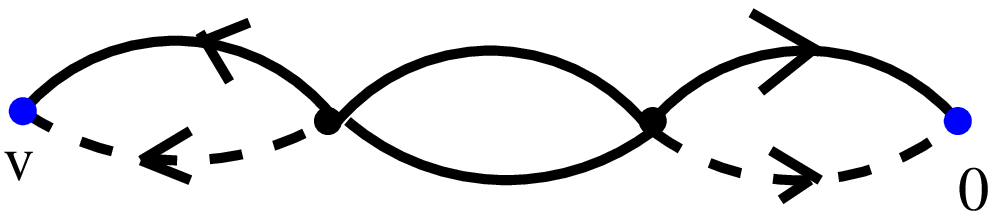}
\end{picture}}
\]
\caption{$\pi^2 \mathcal{K}_4 (v)$}  
\end{center}
\end{figure}

We now consider the graph $\mathcal{K}_4 (v)$ defined by \C{newfour} which is given by
\bea \mathcal{K}_4 (v)= \frac{\tau_2^2}{\pi^2} \int_{wz} \Big\vert \p_z G(z,v)\Big\vert^2 \Big\vert \p_w G(w)\Big\vert^2 G(z,w)^2 \eea
as depicted by figure 11. We rewrite it as
\bea \mathcal{K}_4 (v)= \frac{\tau_2^2}{\pi^2} \int_{wz} \Big\vert \p_z G(z,v)\Big\vert^2 \Big\vert \p_z G(z,w)\Big\vert^2 G(w)^2 \eea
which we now analyze.

In order to simplify $\mathcal{K}_4 (v)$, we start with the auxiliary graph
\bea A_9 (v) = \frac{\tau_2^3}{\pi^3} \int_{wxz}  G(x)^2 \p_z G(z,v) \p_z G(x,z) \overline\p_w G(w,v) \overline\p_w G(w,x) \overline\p_z \p_w G(z,w) \eea
given in figure 8, which leads to
\bea \label{K4}\mathcal{K}_4 (v) &=& 2 E_2^2 + 2 D_4^{(1,1,2)} (v) - \frac{2 D_4^{(1)} (v)}{3} -\frac{2 F_8 (v)}{3} \non \\ && +2 F_{11} (v)  - \Big(F_6 (v) + c.c.\Big) + 2 H_{10}(v),\eea
where the graph $H_{10} (v)$ defined by
\bea H_{10} (v) = \frac{\tau_2^2}{\pi^2} \int_{wxz} G(z,v) G(w,v) \p_x G(x,z) \p_x G(x) \overline\p_x G(w,x) \overline\p_x G(x)\eea
is depicted by figure 7. Thus from \C{relF} we see that $H_{10} (v)$ is the only term which has not simplified in \C{K4}. 
To simplify it, we consider the auxiliary graph
\bea A_{10} (v) = \frac{\tau_2^3}{\pi^3} \int_{wxyz} G(z,v) G(w,v) \p_x G(x,z) \p_x G(x) \overline\p_y G(w,y) \overline\p_y G(y)  \overline\p_x \p_y G(x,y)\eea
given in figure 8, which leads to
\bea \label{H10}&&H_{10} (v) = Q_2 (v) Q_2^* (v) + E_4 +\frac{D_4^{(2)} (v)}{2} - \frac{G_2(v)^2}{2} -2 D_4^{(1,2)} (v) - F_1 (v) \non \\ && +\Big(F_{10} (v) + c.c.\Big) -\frac{1}{2} \Big(\frac{\tau_2}{\pi}  \p_v G_2 (v) \overline\p_v D_3^{(1)}(v) + c.c.\Big)-\Big(\frac{\tau_2}{\pi} \p_v G_2(v) \overline\p_v G_3 (v) + c.c.\Big).\non \\ \eea
Thus on using \C{relF}, \C{relQ} and \C{reldel}, we see that $H_{10} (v)$ has simplified. Hence, adding all the contributions we get that
\bea  \label{k4}&&\mathcal{K}_4 (v) = \frac{1}{4\pi} \Delta_v \Big(D_5^{(2,1,2)} (v) - D_5^{(1,2,2)} (v) - \frac{D_5^{(2)}(v)}{3} - G_2 (v) D_3^{(1)} (v) - 2 G_2 (v) G_3 (v)\Big)\non \\ &&+\frac{1}{4} \Big(\Delta_\tau -16\Big) G_2(v)^2 + 5 E_4 + 2 E_2^2 + 3 D_4^{(1,1,2)} (v) - D_4^{(2)} (v) - 2 D_4^{(1,2)} (v) - 2 G_4 (v) \non \\ &&+ E_2 G_2 (v) + G(v)\Big(D_3^{(1)} (v) - 4 G_3 (v) - G(v) G_2 (v) - E_2  G(v) + 2 E_3\Big).\eea
Thus $\mathcal{K}_4 (v)$ has simplified.

Thus we have obtained non--trivial relations involving the elliptic modular graph functions given by \C{K1}, \C{K2}, \C{3k} and \C{k4}. In fact, there also exist relations between the graphs whose links are given only by the Green function.  
For the modular graphs in figure 4, we have that~\cite{DHoker:2015gmr,DHoker:2015sve,DHoker:2016mwo,Basu:2016kli}
\be \label{relg}C_{1,1,2} = \frac{D_4}{24} +\frac{3 E_4}{4} - \frac{E_2^2}{8},\ee
while for the various graphs in figures 3 and 4, we also have the relation~\cite{Basu:2020pey}
\bea  D_4^{(1,1,2)} (v) - \frac{D_4^{(1)}(v)}{6} +\frac{D_4^{(2)}(v)}{8} - G_4 (v) -\frac{G_2(v)^2}{4} +\frac{1}{2}E_2 G_2 (v) +\frac{E_4}{4} - \frac{E_2^2}{8} =0,\non \\ \eea
which further restrict the relations we have obtained.

\subsection{Relation involving $\mathcal{K}_5 $}

\begin{figure}[ht]
\begin{center}
\[
\mbox{\begin{picture}(250,110)(0,0)
\includegraphics[scale=.55]{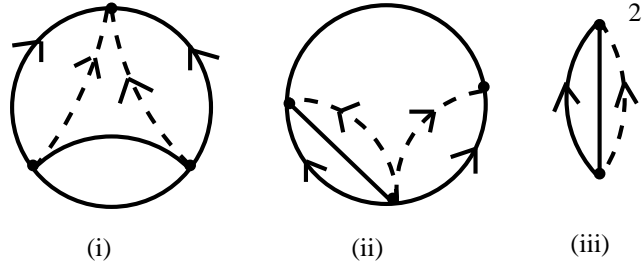}
\end{picture}}
\]
\caption{(i) $\pi^2 \mathcal{K}_5^{(1)} (v)$, (ii) $\pi^2 \mathcal{K}_5^{(2)} (v)$, (iii) $\pi^2 \mathcal{K}_5^{(3)} (v)$}  
\end{center}
\end{figure}

Finally, we consider $\mathcal{K}_5$ defined in \C{newfive} which we write as
\bea \mathcal{K}_5 = \mathcal{K}_5^{(1)} - 2 \mathcal{K}_5^{(2)} + \mathcal{K}_5^{(3)},\eea 
where
\bea \mathcal{K}_5^{(1)} &=& \frac{\tau_2^2}{\pi^2}  \int_{wz}  \Big\vert \p_z G(z)\Big\vert^2  \Big\vert \p_w G(w) \Big\vert^2 G(z,w)^2 ,\non \\ \mathcal{K}_5^{(2)} &=& \frac{\tau_2^2}{\pi^2}  \int_{wz}  \Big\vert \p_z G(z)\Big\vert^2  \Big\vert \p_w G(w)\Big\vert^2 G(z) G(z,w), \non \\ \mathcal{K}_5^{(3)} &=& \Big[ \frac{\tau_2}{\pi}  \int_{z}  \Big\vert \p_z G(z)\Big\vert^2 G(z) \Big]^2  \eea
as depicted by figure 12.

To simplify $\mathcal{K}_5^{(1)}$, we consider the auxiliary graph
\bea A_{11} = \frac{\tau_2^3}{\pi^3} \int_{wxyz}  G(x,y)^2 \p_z G(x,z) \p_z G(y,z) \overline\p_w G(w,x) \overline\p_w G(w,y) \overline\p_z \p_w G(w,z)\eea
given in figure 8,  
leading to
\bea \label{K51}\mathcal{K}_5^{(1)} = \frac{1}{12}\Delta_\tau D_4 + 2 E_2^2 + 2 C_{1,1,2} - \frac{5 D_4}{6}.\eea

We also have that
\be \mathcal{K}_5^{(2)} = \frac{7 D_4}{24} + \frac{C_{1,1,2}}{2} - \frac{E_2^2}{4},\ee
and
\be \mathcal{K}_5^{(3)} = \frac{E_2^2}{4}.\ee
Thus we see that $\mathcal{K}_5$ has simplified, leading to
\bea \mathcal{K}_5 = \frac{1}{12}\Big(\Delta_\tau -\frac{33}{2}\Big)D_4 +\frac{21 E_2^2}{8} +\frac{3 E_4}{4} \eea
on using \C{relg}. 

\subsection{Some consistency checks}

We now perform some non--trivial consistency checks of the results we have obtained. Consider the case where the two points $v$ and 0 are identified in the graphs $\mathcal{K}_i (v)$ ($i=1,\ldots,4$). Thus directly from the figures 6, 9, 10, 11 and 12,   
it follows that
\be \label{relall}\mathcal{K}_1 (0) = \mathcal{K}_2 (0) = \mathcal{K}_3 (0) = \mathcal{K}_4 (0) = \mathcal{K}_5^{(1)}.\ee
We now want to check the consistency of \C{relall}.

First let us consider $\mathcal{K}_1 (0)$, which from \C{K1} yields 
\be \label{match1}\mathcal{K}_1 (0) = \frac{1}{4} \Big(\Delta_\tau -4\Big) E_2^2 + 6 E_4 -\frac{1}{2} D_4 - 2 C_{1,1,2}. \ee
The equality \C{relall} on using \C{K51}, \C{match1} and the constraint \C{relg} yields
the eigenvalue equation
\be \label{poisson}\Big(\Delta_\tau -2\Big) \Big(D_4 - 3 E_2^2\Big) = 36 E_4 - 24 E_2^2,\ee
which is precisely the eigenvalue equation $D_4$ satisfies~\cite{DHoker:2015gmr,DHoker:2016mwo,Basu:2016kli}.

For the remaining graphs, it is not particularly useful to analyze \C{K2}, \C{3k} and \C{k4} directly simply because they have $\Delta_v$ acting on various elliptic graphs, and these expressions have to be simplified before the points $v$ and 0 are identified. Thus for these cases, we directly focus on the graphs that arise at intermediate stages of the calculation where the analysis is considerably simpler as the two unintegrated vertices can be identified. Various identities relevant for our analysis are given in \C{consis1} and \C{consis2} which we use in the expressions below.  

First let us consider $\mathcal{K}_2 (0)$. Using \C{k23}, \C{k25}, \C{k26}, \C{k21}, \C{k22} and \C{k24}, we get that
\be \label{rel2}\mathcal{K}_2 (0) +\mathcal{K}_5^{(1)} =  \frac{1}{2}\Delta_\tau  E_2^2 + 12 C_{1,1,2} -\frac{5D_4}{3}.\ee
Equating the right hand side to $2\mathcal{K}_5^{(1)}$ and using \C{relg} yields \C{poisson} as the consistency condition.

We next consider $\mathcal{K}_3 (0)$ and again directly consider the graphs. Using \C{k31}, \C{k32}, \C{k33} and \C{k34}, we get that
\be \label{rel3}\mathcal{K}_3 (0) = \frac{1}{6} \Big(\Delta_\tau -6\Big) D_4 -\frac{1}{4} \Big(\Delta_\tau -18\Big)E_2^2 -3 E_4 + 2 C_{1,1,2}.\ee
Equating \C{rel3} to $\mathcal{K}_5^{(1)}$ again yields \C{poisson}.

Finally, we consider $\mathcal{K}_4 (0)$. Now \C{K4} and \C{H10} leads to
\be \mathcal{K}_4 (0) = \frac{1}{4}\Big(\Delta_\tau -2\Big) E_2^2 + 3 E_4 + 2 C_{1,1,2} - \frac{2D_4}{3},\ee
which when equated to $\mathcal{K}_5^{(1)}$ yields \C{poisson}. 
Hence we see that the trivial equality of the graphs $\mathcal{K}_i (0)$ ($i=1,\ldots,4$) with $\mathcal{K}_5^{(1)}$ actually yields a non--trivial eigenvalue equation satisfied by the modular graph $D_4$.  

Thus we have shown that the graphs $\mathcal{K}_i (v)$ ($i=1,\ldots,5$) can be expressed in terms of the graphs given in figures 3 and 4, and contain no additional information. The relations are non--linear in the various graphs, and also involve $\Delta_\tau$ and $\Delta_v$ acting on some of the graphs in figures 3 and 4. Note that in each of the equations, the graph involving $\mathcal{K}_i (v)$ has two $\p$ and two $\overline\p$ derivatives and has six links, while each term involving $\Delta_v$ contain graphs with a total of five links. Every other term in each equation contains graphs with a total of four links. This is expected, since using \C{eigen}  we see that for the sake of counting the links, removing a $\p \overline\p$ pair is equivalent to removing a link. 

\appendix

\section{Some elliptic modular graphs and useful identities}

We now consider the expressions for $F_i (v)$, $Q_1$ and $Q_2(v)$ that arise in the main text. Relevant expressions involving them can be expressed in terms of the graphs given in figures 3 and 4 with significantly less effort than the other graphs in the main text.  

We first list the graphs $F_i (v)$ ($i=1,\ldots,11$) that appear in the main text.
They are defined by
\begin{figure}[ht]
\begin{center}
\[
\mbox{\begin{picture}(280,350)(0,0)
\includegraphics[scale=.6]{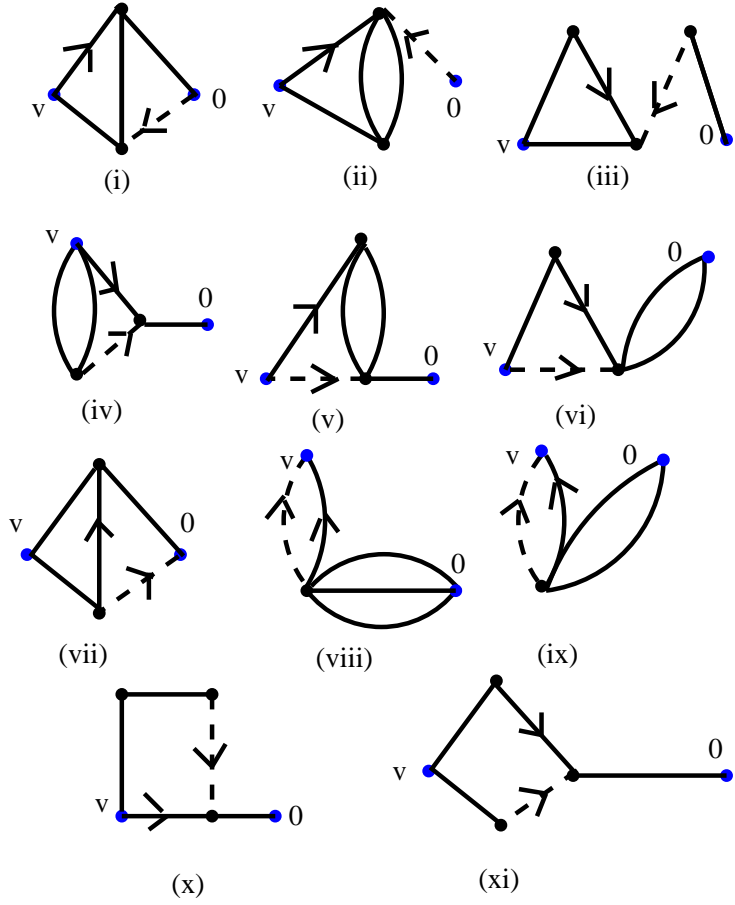}
\end{picture}}
\]
\caption{(i) $\pi F_1  (v)$, (ii) $\pi F_2 (v)$, (iii) $\pi F_3 (v)$, (iv) $\pi F_4 (v)$, (v) $\pi F_5 (v)$, (vi) $\pi F_6 (v)$, (vii) $\pi F_7 (v)$, (viii) $\pi F_8 (v)$, (ix) $\pi F_9 (v)$, (x) $\pi F_{10} (v)$, (xi) $\pi F_{11} (v)$}  
\end{center}
\end{figure}
\bea \label{defF}F_1 (v) &=& \frac{\tau_2}{\pi} \int_{wz} G(z) G(w,v) G(w,z)\p_z G(z,v) \overline\p_w G(w), \non \\ F_2 (v) &=& \frac{\tau_2}{\pi} \int_{wz} G(w,v) G(w,z)^2 \p_z G(z,v) \overline\p_z G(z), \non \\ F_3 (v) &=& \frac{\tau_2}{\pi} \int_{wxz} G(w) G(x,v) G(z,v) \p_z G(x,z) \overline\p_z G(z,w), \non \\ F_4 (v) &=& \frac{\tau_2}{\pi} \int_{wz}G(z) G(w,v)^2 \p_z G(v,z) \overline\p_z G(w,z), \non \\ F_5 (v) &=& \frac{\tau_2}{\pi} \int_{wz} G(w) G(w,z)^2 \p_z G(z,v) \overline\p_w G(w,v), \non \\ F_6 (v) &=& \frac{\tau_2}{\pi} \int_{wz} G(z)^2 G(w,v) \p_z G(z,w) \overline\p_z G(z,v), \non \eea
\bea
F_7 (v) &=& \frac{\tau_2}{\pi} \int_{wz} G(z) G(z,v) G(w,v) \p_w G(w,z) \overline\p_w G(w), \non \\ F_8 (v) &=& \frac{\tau_2}{\pi} \int_{z} G(z)^3 \p_z G(z,v) \overline\p_z G(z,v), \non \\ F_9 (v) &=& \frac{\tau_2}{\pi} \int_{z} G(z)^2 \p_z G(z,v) \overline\p_z G(z,v), \non \\ F_{10} (v) &=& \frac{\tau_2}{\pi} \int_{wxz} G(z) G(x,v) G(x,w) \p_z G(z,v) \overline\p_w G(w,z),  \non \\   F_{11} (v) &=& \frac{\tau_2}{\pi} \int_{wxz} G(z) G(w,v) G(x,v) \p_z G(w,z) \overline\p_z G(z,x),\eea
as depicted by figure 13.

We also define the graphs $Q_1$ and $Q_2 (v)$ by\footnote{Note that $Q_2 (0) = Q_1$.}
 \bea \label{defQ}Q_1 &=& \frac{\tau_2}{\pi} \int_{z_1 z_2 z_3}  \p_{z_1} G(z_1,z_2) \p_{z_1} G(z_1,z_3) G(z_2,z_3) , \non \\ Q_2 (v) &=& \frac{\tau_2}{\pi} \int_{z_1 z_2}  \p_{z_1} G(z_1,v) \p_{z_1} G(z_1,z_2) G(z_2)\eea
which transform as
\be Q_1 \rightarrow (c\tau+d)(c\overline\tau +d)^{-1} Q_1, \quad  Q_2 (v) \rightarrow (c\tau+d)(c\overline\tau +d)^{-1} Q_2 (v)\ee
under $SL(2,\mathbb{Z})$ transformations.

The graphs \C{defF} satisfy useful identities given by 
\bea \label{relF}F_1 (v) &=& - G(v) D_3^{(1)} (v) + D_4^{(1,2)} (v) +\frac{D_4^{(2)}(v)}{2} -\frac{ G_2(v)^2}{2} -\frac{1}{8\pi} \Delta_v D_5^{(2,1,2)} (v), \non \\ F_2 (v) + c.c. &=& G(v) D_3^{(1)} (v) -  C_{1,1,2} +  G(v) D_3 -  D_4^{(1,1,2)} (v) -  D_4^{(1)} (v) + E_2 G_2 (v),\non \\ F_3 (v) + c.c. &=&  D_4^{(1,2)} (v) + D_4^{(1,1,2)} (v) + G_4 (v) - E_2 G_2 (v), \non \\ F_4 (v) + c.c.  &=&  D_4^{(1)} (v) + D_3 G(v) - D_4^{(1,1,2)} (v)  - E_2 G_2 (v) + C_{1,1,2} - G(v) D_3^{(1)} (v),\non\\ F_5 (v) + c.c. &=&  C_{1,1,2} - G(v) D_3^{(1)} (v) + D_4^{(1)} (v) - E_2 G_2 (v) + D_3 G(v) - D_4^{(1,1,2)} (v),\non \\ F_6 (v) + c.c. &=&  E_2 G(v)^2 - D_4^{(1,1,2)} (v) + D_4^{(2)} (v)+\frac{1}{4\pi} \Delta_v D_5^{(1,2,2)} (v),\non \\ F_7 (v) &=&\frac{D_4^{(2)} (v)}{2} -\frac{ G_2(v)^2}{2}, \non \\ F_8 (v) &=& - D_4^{(1)}(v)+ \frac{1}{8\pi} \Delta_v D_5^{(2)} (v),  \non \\ F_9 (v) &=& - D_3^{(1)}(v)+ \frac{1}{8\pi} \Delta_v D_4^{(2)} (v), \non \\ F_{10}(v) + c.c. &=&  E_4-  G(v) G_3 (v) +  D_4^{(1,2)} (v) + E_3 G(v) - G_4 (v), \non \\   F_{11} (v) &=&  D_4^{(1,2)} (v) +\frac{E_4}{2}  - \frac{G_2(v)^2}{2} \eea 
which we often use. When the vertices at $v$ and 0 are identified, these elliptic modular graphs reduce to modular graphs, hence satisfying several identities, of which the identities  
\bea \label{consis1} &&F_1 (0) = \frac{F_7 (0)}{2} = \frac{1}{4} \Big(D_4 - E_2^2\Big), \quad F_2 (0) = \frac{1}{2} \Big(E_2^2 - D_4\Big)-C_{1,1,2}, \non \\ && F_3 (0) = F_{11} (0) = C_{1,1,2} + \frac{1}{2} \Big(E_4 - E_2^2\Big), \quad F_6 (0) = -\frac{4F_8 (0)}{3}=\frac{D_4}{3}, \quad F_{10} (0) = \frac{C_{1,1,2}}{2}\non \\ \eea
are used in the main text.

The $SL(2,\mathbb{Z})$ invariant Laplacian $\Delta_\tau$ can be expressed as  
\be \label{Belt}\Delta_\tau = \overline\p_\mu\p_\mu\ee
where $\p_\mu$ is the variation under complex structure deformations and $\mu$ is the Beltrami differential. 
 In order to obtain the action of $\Delta_\tau$ on various graphs using \C{Belt}, we use the relations~\cite{Verlinde:1986kw,DHoker:1988pdl,DHoker:2015gmr,Basu:2015ayg}
\be \label{Beltrami}\p_\mu G(z_1,z_2) = -\frac{\tau_2}{\pi}\int_{z}\p_z G(z,z_1) \p_z G(z,z_2),\ee
and
\be \overline\p_\mu\p_\mu G(z_1,z_2)=0.\ee
Along with the eigenvalue equations
\be \Delta_\tau E_k = k(k-1) E_k,\quad \Delta_\tau G_k (v) = k(k-1) G_k (v),\ee
for $k \geq 2$, we see that the $SL(2,\mathbb{Z})$ invariant combinations of the the graphs appearing in \C{defQ} yield the relations
\bea \label{relQ}Q_1 Q_1^* &=& \frac{1}{8} \Big(\Delta_\tau -4\Big) E_2^2, \non \\ Q_2  (v) Q_2^* (v) &=& \frac{1}{8} \Big(\Delta_\tau -4\Big) G_2 (v)^2, \non \\ Q_1 Q_2^* (v)+ c.c. &=& \frac{1}{4} \Big(\Delta_\tau -4\Big) \Big( E_2 G_2 (v) \Big).\eea
On identifying $v$ and 0 in $Q_2 (v)$, these yield the relations
\bea \label{consis2}Q_1 Q_1^* = Q_2 (0) Q_2^*(0) = \frac{1}{2}\Big(Q_1^* Q_2 (0) + c.c.\Big)= \frac{1}{8}\Big(\Delta_\tau -4\Big) E_2^2 \eea
which are used in our analysis.

Finally, we make use of the identities 
\bea \label{reldel}\frac{\tau_2}{\pi} \p_v G_2 (v) \overline\p_v G_3 (v) + c.c. &=& \frac{1}{4\pi} \Delta_v \Big( G_2 (v) G_3 (v)\Big) + G_2 (v)^2 + G(v) G_3 (v), \non \\ \frac{\tau_2}{\pi} \p_v D_3^{(1)}(v) \overline\p_v G_2 (v) + c.c. &=& \frac{1}{4\pi}\Delta_v \Big(D_3^{(1)} (v) G_2 (v)\Big) + G_2 (v) G(v)^2 \non \\ && - E_2 G_2 (v)+ G(v) D_3^{(1)} (v), \non \\ \frac{\tau_2}{\pi} \p_v G_2 (v) \overline\p_v G_2 (v) &=& \frac{1}{8\pi} \Delta_v G_2(v)^2 + G(v) G_2 (v)\eea
to express the graphs with factors of $\p_v$ and $\overline\p_v$ along the links in terms of graphs with no derivatives of the Green functions in the links, at the cost of a contribution that has $\Delta_v$ acting on it.     

\providecommand{\href}[2]{#2}\begingroup\raggedright\endgroup


\end{document}